# Beyond the Inverted Index

Zhi-Hong Deng

Key Laboratory of Machine Perception (Ministry of Education), School of Electronics
Engineering and Computer Science, Peking University, Beijing 100871, China
E-mail addresses: zhdeng@pku.edu.cn

**Abstract:** In this paper, a new data structure named group-list is proposed. The group-list is as simple as the inverted index. However, the group-list divides document identifiers in an inverted index into groups, which makes it more efficient when it is used to perform the intersection or union operation on document identifiers. The experimental results on a synthetic dataset show that the group-list outperforms the inverted index.

**Keywords**: data structure; the inverted index; the group-list

## 1. Introduction

The inverted index is the data structure at the core of most large-scale search systems for text, (semi-) structured data, and graphs, with web search engines, XML and RDF databases, and graph search engines in social networks [Ottaviano and Venturini 2014]. How to make the inverted index more suitable for corpora with huge size is always the key issue since the 1950s. Most of the previous works focus on the trade-off between space occupancy and decompression speed [Lemire and Boytsov 2013; Moffat and Stuiver 2000; Salomon 2007; Stepanov et al. 2011; Yan et al. 2009].

In recent years, we have proposed some kinds of data structure similar to the inverted index, named Node-list [Deng and Wang 2010], N-list [Deng et al. 2012], Nodeset [Deng and Lv 2014], and DiffNodeset [Deng 2016], to promote the efficiency of frequent itemset mining. Many studies [Deng and Lv 2015; Vo et al. 2016; Vo et al. 2017; Aryabarzan et al. 2018; Huynh et al. 2019; Han et al. 2019] show that these structures are very efficient for mining frequent itemsets.

However, these kinds of structure store only the summary information about frequent items, which can not be used to search or query infrequent items directly. In this paper, we extend the Node-list structure to enable the ability of searching both frequent items and infrequent items, just like the inverted index. Since we focus on an index in place of the inverted index for information retrieval in this paper, we use term instead of item as the basic element.

The major contributions of this paper are as follows:
(1) A novel data structure, named group-list, is proposed. The group-list of a term maintains the identifiers of documents containing the term in terms of group, which makes it more efficient for searching documents containing a set of terms when compared with the inverted index.
(2) Some basic experiments over a synthetic dataset were conducted to compare the group-list with the inverted index. Experimental results show that the group-list outperforms the inverted index, especially for searching a set of frequent terms.

The rest of this paper is organized as follows. Section 2 presents the group-list structure and relevant concepts and corresponding construction algorithms. Section 3 develops the methods of query processing based on the group-list structure. Experiment results are shown in Section 4 and conclusions and future work are given in Section 5.

## 2. Notation, concepts, and construction algorithms

Let $I = \{i_1, i_2, \ldots, i_M\}$ be the universal set of terms, $D = \{Id_1, Id_2, \ldots, Id_N\}$, where $Id_k$ is the document identifier of $k$-th document, be a collection of documents. In this paper, each document is regarded as a set of term for simplicity.

**Definition 1.** The count of term $i$ in $D$, denoted as $C_i$, is the number of documents where term $i$ occurs.

**Definition 2.** Given percentage threshold $\zeta$, term $i$ is called a frequent term if and only if its count is not less than $\zeta \times |D|$, where $|D|$ is the number of documents in $D$.

Before defining the group-list structure, we first introduce prefix tree, which is the basis of group-list. The prefix tree is a compact structure that maintains sufficient information about terms in document set $D$.

**Definition 3.** Given document set $D$ and percentage threshold $\zeta$, the prefix tree (P-tree) of $D$ is defined below.
  (1) It is made up of one root labeled as "Root", a set of term prefix subtrees as the children of the root.
  (2) Each node consists of five fields: *label*, *n_code*, *Id_set*, and *parent-link*. The description of each field is listed as follows.
     (2.1) Field *label* registers all terms that this node represents.
     (2.2) Field *n_code*, registers a unique identifier that represents the node.
     (2.3) Field *Id_set* contains the identifiers of all documents that contain a term registering in the node. *Id_set* of a node consists of a set of records. Each record includes two fields: *i_id* and *did_set*. *i_id* indicates a term registering in the node, and *did_set* contains the identifiers of all documents that contain the term indicated by *i_id*.
     (2.4) Field *parent-link* points to the parent of the node.

Based on definition 3, the P-tree of a document set D can be constructed by algorithm 1.

**Algorithm 1: P-tree Construction**

**Input:** A collection of documents $D$ and a percentage threshold $\zeta$.
**Output:** $PT$, a P-tree.
1: Scan $D$ once to compute the count of each term, and sort each document $d$ in $D$ according to the count descending order.
2: initialize $PT$ with root $Root$;
3: **For** each document $d$ in $D$ **do**
4:     $i \leftarrow d.first\text{-}term$;
5:     $d \leftarrow d - \{i\}$;
6:     Call Insert_Tree($i, d, Root$);
7: **Return** $PT$;

**Function Insert_Tree($t, Str, Nd$)**

1: **If** $Nd$ has a child $N$ such that $t \in N.label$ **then**
2:     Find the record which $i\_id$ is $t$ in $N.Id\_set$. Let the record be $Rec[t]$;
3:     $Rec[t].did\_set \leftarrow Rec[t].did\_set \cup \{Str.Id\}$;
4:     $Next\_Node \leftarrow N$;
5: **else**
6:     **If** ($t$ is frequent term) **or** ($t$ is the first infrequent term)  // $t$ is frequent term if its count is *is not less than* $\zeta \times |D|$
7:        create a new node $N$;
8:        $N.label \leftarrow \{t\}$;
9:        $N.parent\text{-}link \leftarrow Nd$;
10:       $Temp\_Rec.i\_id \leftarrow t$;
11:       $Temp\_Rec.did\_set \leftarrow \{Str.Id\}$;  // the identifier of the current document
12:       $N.Id\_set \leftarrow \{Temp\_Rec\}$;
13:       $Next\_Node \leftarrow N$;
14:     **else**
15:       $Nd.label \leftarrow Nd.label \cup \{t\}$;
16:       $Temp\_Rec.i\_id \leftarrow t$;
17:       $Temp\_Rec.did\_set \leftarrow \{Str.Id\}$;
18:       $Nd.Id\_set \leftarrow \{Temp\_Rec\}$;
19:       $Next\_Node \leftarrow Nd$;
20: **If** $Str \neq \varnothing$ **then**
21:     $t_{next} \leftarrow Str.first\text{-}term$;
22:     $Str \leftarrow Str - \{t_{next}\}$;
23:     **Call Insert_Tree($t_{next}, Str, Next\_Node$)**;

It needs two scan of document set $D$ to construct its P-tree. Algorithm 1 shows the details. Initially, a tree with root $Root$ is first created as shown by Line 1. Subsequently, Line 2 to 6 construct the P-tree by processing documents one by one. Each document is inserted into the tree by calling Insert_Tree($t, Str, Nd$).

The function Insert_Tree($t, Str, Nd$) is performed as follows. If $Nd$ has a child node $N$ such that $N.label$ contains $t$, then the identifier of the current document is appended to $Tri[t].did\_set$, which is the set of the identifier of documents in which $t$ occurs. Otherwise, if $t$ is frequent term or $t$ is the first infrequent term, a new node $N$ with $t$ as initial label is created. Tuple $Temp\_Tri$ containing information about $t$ is then inserted into the $Id\_set$ of node $N$, $N.Id\_set$. If none of the aforementioned conditions is satisfied, $t$ is added to $N.label$, the label of current node $Nd$, and Tuple $Temp\_Tri$ containing information about $t$ is then inserted into $Nd.Id\_set$. Finally, if current document (term sequence) is not null, its first term is taken out and call function Insert_Tree() recursively.

According to Algorithm 1, all frequent terms register in the non-leaf nodes (except root), and the label of a non-leaf node (except root) contains only one frequent term. All infrequent terms register in the leaf nodes, and the label of a leaf node contains one or more infrequent terms. The number of frequent terms affects the efficiency of terms searching, which will be presented and discussed in the experiment section. The percentage threshold $\zeta$ is used to control the number of frequent terms.

After constructing the P-tree, it is easy to obtain the group-list of each term. Algorithm 2 describes the details.

As mentioned above, the label of a non-leaf node contains only one term and thus one record in its $Id\_set$ field. Conversely, the label of a leaf-node contains one or more term and thus one more records in its $Id\_set$ field. Therefore, we should make use of these characteristics in implementing these algorithms for better efficiency.

**Algorithm 2: Group-list Generation**

**Input:** *PT*, a P-tree.
**Output:** the group-list of each term.
  1: Scan *PT* once with preorder traversal to obtain the *pre-order* of each node, and once again with postorder traversal to obtain the *post-order* of each node.
  2: for each term, initialize its group-list with *null*;
  3: **While** scan *D* by preorder traversal **do**
  4:   *Temp_Node* ← current visiting node;
  5:   **For** each *i* in *Temp_Node.label* **do**
  6:     Find the record which *i_id* is *i* in *Temp_Node.Id_set*. Let the record be *Rec*[*i*];
  7:     Append{(<*Temp_Node. pre-order, Temp_Node. post-order*>: *Rec*[*i*].*did_set*)}to the group-list of *i*;

According to algorithm 2, the group-list of a term is a set of tuples. Each of these tuples contains the pre-order and post-order of a node registered by the term, and the set of identifiers of documents containing the term, which is collected in the node. **Each tuple has two fields: *pp_order* and *did_set*. Field *pp_order* contains two subfields: *pre-order* and *post-order*.** Field *pp_order* contains the pre-order and post-order of a node while field *did_set* contains a set of identifiers of documents. It should be noted that the tuples in a group-list is sorted by the *pre-order* ascending order according to algorithm 2. This characteristic is very helpful for query processing.

For better understanding the aforementioned algorithms and concepts, let's examine an example as follows.
**Example 1**. Table 1 presents a collection of documents, *CD*. Each row of Table 1 stands for a document. The left column contains the document identifier and the right column contains the set of terms that occur in a document. In this example, we set $\zeta = 50\%$.

Table 1. A collection of documents, *CD*

| Document identifier | terms |
|---|---|
| 1 | a, c |
| 2 | b, c, e, g, h |
| 3 | a, b, c, e, h |
| 4 | b, e |
| 5 | a, c, d, f, i |
| 6 | b, c, e, h |
| 7 | b, e, i |
| 8 | a, b, c, e, f, h |
| 9 | a, b, c, d, e, f |
| 10 | d, f |

After scanning *CD*, we obtain that the count of *a, b, c, d, e, f, g, h,* and *i* are 5, 7, 7, 3, 7, 4, 1, 4, and 2 respectively. By definition 2, *a, b, c,* and *e* are frequent terms while other 5 terms are infrequent. According to the count descending order, the sorted term set, $l_S$, is {*b, c, e, a, f, h, d, i, g*}. Therefore, the P-tree is actually constructed over the sorted document set as Table 2. Figure 1 to 4 present the construction procedure of the P-tree according to algorithm 1. Figure 5 shows the P-tree with each node encoded < *pre-order, post-order*>, which is obtained by scanning the tree with preorder traversal and postorder traversal. Figure 6 presents the group-list of each term.

According to algorithm 1, the group-list of a term contains the identifiers of all documents in which the term occurs. In fact, the union of all *did_set* in the group-list is just the set of the identifiers of all documents. For example, examine *c*→{(<1,2>:{1,5}), (<5,7>:{2,3,6,8,9})} in Figure 6. {1,5} ∪ {2,3,6,8,9} is {1, 2, 3, 5, 6, 8, 9}. As shown by Table 2, only documents with identifier 1, 2, 3, 5, 6, 8, and 9 contain *c*. Figure 6 shows the group-list (left) and the inverted index (right) of each term. Obviously, the document identifiers in the group-list and the inverted index of a term are the same.

Table 2. The sorted document set, *SCD*, of *CD*

| Document identifier | terms |
|---|---|
| 1 | c, a |
| 2 | b, c, e, h, g |
| 3 | b, c, e, a, h |
| 4 | b, e |
| 5 | c, a, f, d, i |

| 6 | b, c, e, h |
| 7 | b, e, i |
| 8 | b, c, e, a, f, h |
| 9 | b, c, e, a, f, d |
| 10 | f, d |

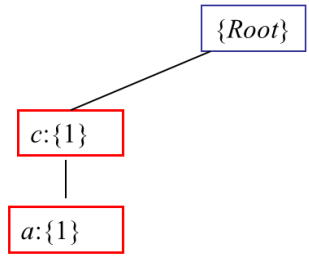

Figure 1: the F-tree after processing Document 1, {c, a}

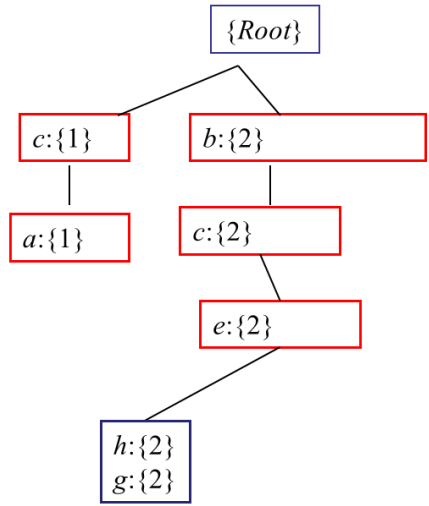

Figure 2: the F-tree after processing Document 2, {b, c, e, h, g}

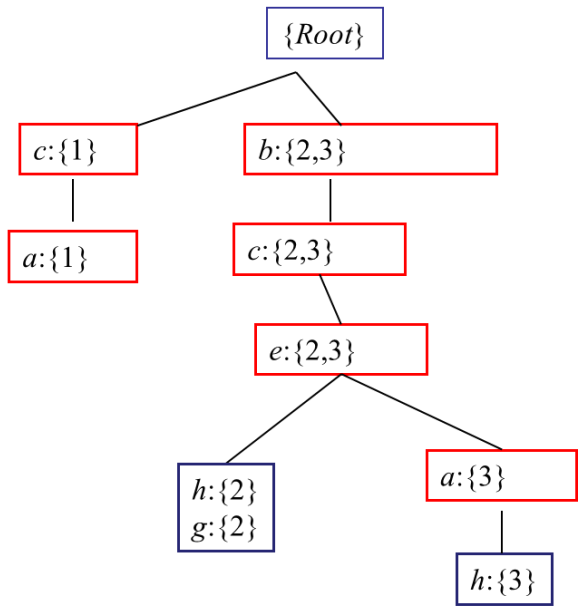

Figure 3: the F-tree after processing Document 3, {b, c, e, a, h}

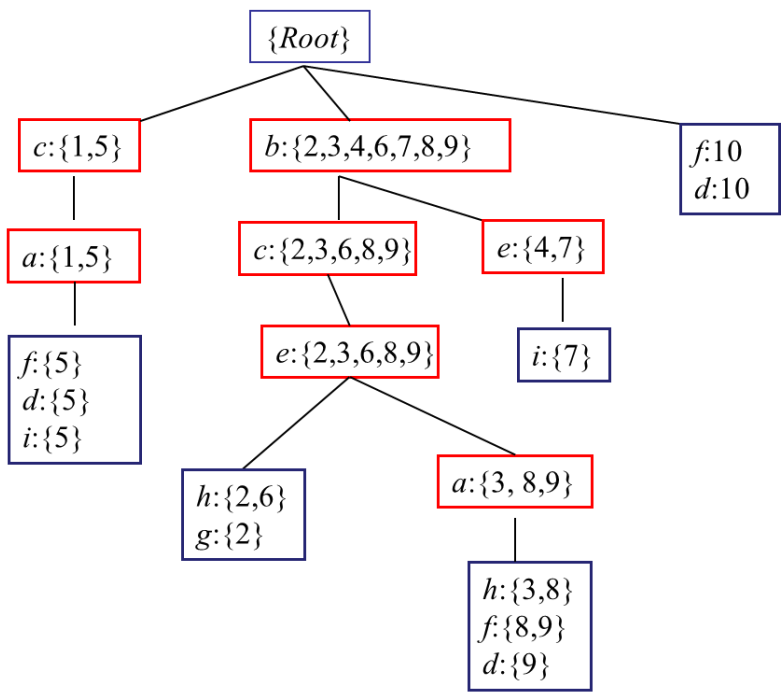

Figure 4: the final F-tree after processing Document 10, {f, d}

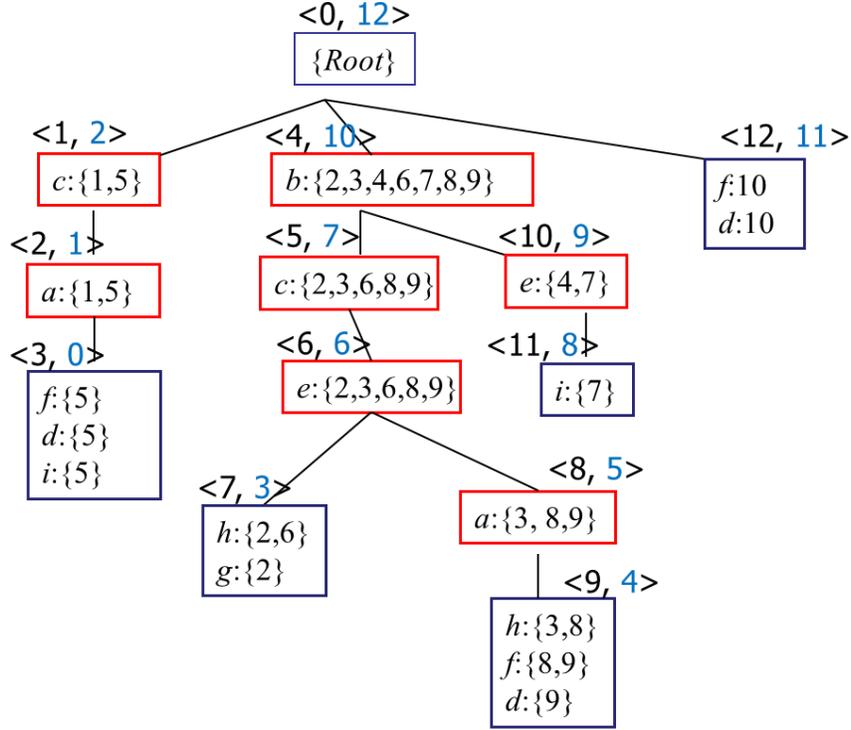

Figure 5: the final F-tree with each node encoded < *pre-order, post-order*>

| | |
|---|---|
| b→{(<4,10>:{2,3,4,6,7,8,9})} | b:{2,3,4,6,7,8,9} |
| c→{(<1,2>:{1,5}), (<5,7>:{2,3,6,8,9})} | c:{1,2,3,5,6,8,9} |
| e→{(<6,6>:{2,3,6,8,9}), (<10,9>:{4,7})} | e:{2,3,4,6,7,8,9} |
| a→{(<2,1>:{1,5}), (<8,5>:{3,8,9})} | a:{1,3,5,8,9} |
| f→{(<3,0>:{5}), (<9,4>:{8,9}), (<12,11>:{10})} | f:{5, 8, 9,10} |
| h→{(<7,3>:{2,6}), (<9,4>:{3,8})} | h:{2,3,6, 8} |
| d→{(<3,0>:{5}), (<9,4>:{9}), (<12,11>:{10})} | d:{5,9,10} |
| i→{(<3,0>:{5}), (<11,8>:{7})} | i:{5,7} |
| g→{(<7,3>:{2})} | g:{2} |

Figure 6: the group-list (**left**) and the inverted index (**right**) of each term

In fact, the group-list is the same as the node-list [Deng and Wang 2010]. The only difference between them is that the group-list indexes both frequent and infrequent terms (items) while the node-list only indexes only frequent terms (items). Therefore, the group-list naturally possesses the properties possessed by the node-list. Based on the proof methods provided in [Deng and Wang 2010], we have the property:

**Property 1**: Given a set of term $\{t_1, t_2, \ldots, t_n\}$, assume that the group-list of $t_i$ is group-list$[t_i]$, the group-list of termset $t_1t_2 \ldots t_n$ by intersecting all group-list$[t_i]$ ($1 \leq i \leq n$). The identifiers of all documents that contain $t_1, t_2, \ldots, t_n$ are the union of all *did_sets* in the group-list of $t_1t_2 \ldots t_n$.

For the intersecting operation, please refer to Definition 7 (the Node-list of a k-pattern) in [Deng and Wang 2010]. Figure 7 present an example of how to construction the group-list of termset *ba*. In fact, the group-list of *ba* is the tuples in the group-list of *a*. the node corresponding to the ***pp_order*** of each of these tuples must be a descendant of the node corresponding to the ***pp_order*** of some tuple in the group-list of *b*. the descendant relationship can be easy judged by *pre-order* and *post-order,* as shown in [Deng and Wang 2010] and following algorithm 3. Property 1 provides an efficient method for query processing.

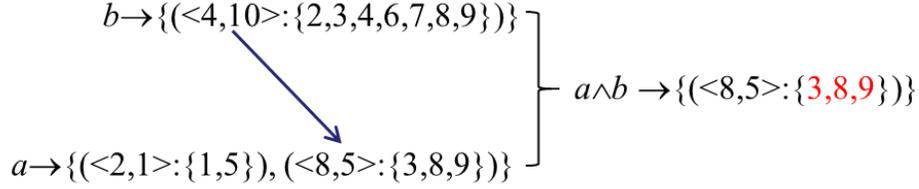

Figure 7: An example of construction the group-list of termset *ba*

## 3. Query processing

As mentioned in [Ottaviano and Venturini 2014]. The query processing of information retrieval can be described as follows. Given a term query as a set of terms, the basic query operation are the Boolean conjunctive (AND) and disjunctive (OR) queries, retrieving the documents that contain respectively all the terms or at least one of them. In this paper, we only consider two types of queries. For simplicity, we denote the Boolean conjunctive (AND) query as BAND query, and Boolean disjunctive (OR) query as BOR query. For both BAND and BOR query, the query processing based on the group-list is as simple as (or very similar to) that based on the inverted index.

Given a set of term $\{t_1, t_2, \ldots, t_K\}$, the retrieval results of the BAND query is to obtain all document that contain these terms. For the sake of discussion, we assume that the count of $t_i$ is not less than he count of $t_j$ if $i < j$. That is, $\{t_1, t_2, \ldots, t_k\}$ is a sorted list in terms of the count descending order. Furthermore, we assume that $t_1, t_2, \ldots,$ and $t_f$ are frequent terms while $t_{f+1}, t_{f+2}, \ldots,$ and $t_K$ are infrequent terms. As mentioned in the last paragraph of Section 2, the document identifiers in the group-list of a term are the same as in the inverted index of the term. Therefore, we just need to find these common document identifiers in the group-lists of all query terms. To this end, we propose the following algorithm.

---

**Algorithm 3:** BAND query processing

**Input:** a query, $\{t_1, t_2, \ldots, t_K\}$.
**Output:** $Res\_did$, the set of identifiers of all documents that contain the $K$ terms.
  // Line 1 to 7 obtains the group-list of termset $t_{f+1}t_{f+2}\ldots t_K$ by joining the group-lists of $t_{f+1}, t_{f+2}, \ldots,$ and $t_K$
  // $GL_{f+1:\,K}$ stand for the group-list of termset $t_{f+1}t_{f+2}\ldots t_K$. It holds the identifiers of all documents contains $t_{f+1}, t_{f+2}, \ldots,$ and $t_K$
1:  $GL_{f+1:\,K} \leftarrow$ group-list$[t_{f+1}]$ ; // group-list$[t_{f+1}]$ is the group-list of $t_{f+1}$
2:  **For** $i = 2$ to $K$ **do**
3:    **For** each tuple $tp$ in $GL_{f+1:\,K}$ **do**
4:      **If** $\exists$ tuple $tp^* \in$ group-list$[t_{f+i}]$, $tp^*.pp\_order.pre\text{-}order = tp.pp\_order.pre\text{-}order$, **then**
5:        $tp.did\_set \leftarrow tp.did\_set \cap tp^*.did\_set$;
6:      **Else**
7:        delete $tp$ from $GL_{f+1:\,K}$;
  // Line 8 to 15 obtains the group-list of termset $t_1t_2\ldots t_f$ by joining the group-lists of $t_1, t_2, \ldots,$ and $t_f$
  // $GL_{1:\,f}$ stand for the group-list of termset $t_1t_2\ldots t_f$. It holds the identifiers of all documents contains $t_1, t_2, \ldots,$ and $t_f$
8:  $GL_{1:\,f} \leftarrow$ group-list$[t_1]$ ; // group-list$[t_1]$ is the group-list of $t_1$
9:  **For** $i = 2$ to $f$ **do**
10:   **For** each tuple $tp$ in $GL_{1:\,f}$ **do**
11:     **If** $\exists$ tuple $tp^* \in GL_i$, $(tp.pp\_order.pre\text{-}order < tp^*.pp\_order.pre\text{-}order) \wedge (tp.pp\_order.post\text{-}order > tp^*.pp\_order.post\text{-}order)$ **then**
       // replace the current value of $tp$ with the value of $tp^*$
12:       $tp.pp\_order \leftarrow tp^*.pp\_order$; // assign the value of $pp\_order$ of $tp^*$ to the $pp\_order$ of $tp$
13:       $tp.did\_set \leftarrow tp^*.did\_set$; // assign the value of $did\_set$ of $tp^*$ to the $did\_set$ of $tp$
14:     **Else**
15:       delete $tp$ from $GL_{1:\,f}$;
// generate $Res\_did$, the set of identifiers of all documents that contain the $K$ terms.
16: $Res\_did \leftarrow \varnothing$;
17: For each tuple $tp$ in $GL_{f+1:\,K}$ **do**
18:   **If** $\exists$ tuple $tp^* \in GL_{1:\,f}$, $(tp^*.pp\_order.pre\text{-}order < tp.pp\_order.pre\text{-}order) \wedge (tp^*.pp\_order.post\text{-}order > tp.pp\_order.post\text{-}order)$ **then**
19:     $Res\_did \leftarrow Res\_did \cup tp.did\_set$;

---

For Line 11 and Line 18, the Algorithm 2 (code-intersection) in [Deng and Wang 2010] provides a linear method to implement the operation. Please refer to [Deng and Wang 2010] for details.

For BOR query, the processing is almost the same as that of BAND query (**Algorithm 3**) except that union is considered instead of intersection.

It should be noted that algorithm 4 provides a simple way to deal with the intersection of the group-lists of infrequent terms. Since all infrequent terms register in leaf-nodes, we can use the *pre-order* to judge whether these infrequent terms are in the same leaf-node.

## 4. Experimental Evaluation

In this section, we present a performance comparison of the group-list with the inverted index on a synthetic datasets. The experiments were performed on a PC server with 16G memory and 2GHZ Intel processor. All codes were implemented by C# and ran on X64 windows server 2003 system

We generated a large synthetic database by IBM Quest Synthetic Data Generator (https://github.com/zakimjz/IBMGenerator) . The synthetic database named Syn_data is used in our experiments. To generate Syn_data, the average transaction size, the number of transactions, and number of different items are set to 60, 1000K, and 1K respectively.

We generated three types of queries: (1) queries randomly selected only from frequent terms; (2) queries randomly selected from all terms; (3) queries randomly selected only from infrequent terms. For each type of queries, we generated 200 queries with length of 2, 4, and 6 respectively. For simplicity, we denote the 9 groups of queries as FQ2, FQ4, FQ6, MQ2, MQ4, MQ6, IQ2, IQ4, and IQ6 where FQ, MQ, IQ stand for queries randomly selected only from frequent terms, queries randomly selected from all terms, and queries randomly selected only from infrequent terms respectively. Each group contains 200 queries.

We conducted two experiments over the Syn_data by setting percentage threshold $\zeta$ to be 81% and 90%. When $\zeta$ is 81%, the number of frequent terms is 194. The group-lists of frequent terms contain 1955 tuples and each tuples contains 90 document identifiers on average. The inverted lists of frequent terms contain 904,573 documents identifiers on average. The size of all group-lists is 2,086,760,13 bytes. When $\zeta$ is 90%, the number of frequent terms is 96. The group-lists of frequent terms contain 384 tuples and each tuples contains 237 document identifiers on average. The inverted lists of frequent terms contain 952,312 documents identifiers on average. The size of all group-lists is 2,057,483,392 bytes.

It should be noted that the size of inverted index is 2,009,037,280 bytes. The size of the group-list is as almost the same as that of the inverted index. In fact, compared with the inverted index, the extra size of the group-list is the room for holding the *pre-order* and *post-order*, which is negligible in terms of the size of the inverted index.

Table 3 and 4 present the results of running time for query processing. Obviously, the group-list performs much better than the inverted index on the queries that consist of terms selected from frequent terms or all terms randomly. For the queries consisting of infrequent terms, the inverted index performs a little better than the group-list. The reason can be explained as follows. For frequent terms, they share a lot of common document identifiers in the *did_set* fields of the tuples of their group-list. Therefore, we just compare the *pre-order* and *post-order* once to get a lot of common document identifiers. This avoids the matching of each documents identifier in two inverted index, which is especially efficient when the size of inverted index is huge. As for queries consisting of infrequent terms, the shared document identifiers in the *did_set* fields of the tuples are small and sparse. Therefore, the advantage that the common identifiers are shared in group-lists is diminished.

|  | FQ2 | FQ4 | FQ6 | MQ2 | MQ4 | MQ6 | IQ2 | IQ4 | IQ6 |
|---|---|---|---|---|---|---|---|---|---|
| Inverted index | 12.65 | 34.66 | 53.55 | 12.65 | 33.85 | 56.04 | 4.93 | 7.97 | 9.1 |
| Group-list | 6.49 | 16.47 | 23.98 | 6.92 | 16.75 | 24.07 | 7.37 | 10.85 | 12.01 |

Table 3. Running time (Sec.) comparison when $\zeta$ is 81%

|  | FQ2 | FQ4 | FQ6 | MQ2 | MQ4 | MQ6 | IQ2 | IQ4 | IQ6 |
|---|---|---|---|---|---|---|---|---|---|
| Inverted index | 16.01 | 46.93 | 74.82 | 16.47 | 47.33 | 77.57 | 7.35 | 11.98 | 13.07 |
| Group-list | 4.83 | 9.68 | 13.76 | 3.58 | 8.88 | 13.48 | 8.96 | 11.39 | 13.23 |

Table 4. Running time (Sec.) comparison when $\zeta$ is 90%

## 5. Conclusion

In this paper, we proposed a data structure, group-list, to promote the efficiency of information retrieval. The experimental results show that the group-list outperforms the inverted index, the classic index in information retrieval.

In the future, there are a lot of topics needing to explore. First, there are many methods and technologies based on the inverted index. It is very interesting work to adopt the group-list instead of the inverted index in these methods and technologies for better performance. Second, it is also interesting to design similar indexes based on N-list [Deng et al. 2012], Nodeset [Deng and Lv 2014], and DiffNodeset [Deng 2016]. Finally, since the group-list divides the data into different group, it is naturally suitable to be processed parellelly. Therefore, the parallel/distributed implementation of group-list is also a very interesting work.